\documentclass{article}

\usepackage{graphicx}
\usepackage{ subfig}
\usepackage{color}
\usepackage{bm}
\usepackage{mathrsfs}
\usepackage{ulem}
\usepackage{multirow}
\usepackage{pdflscape}
\usepackage{rotating}
\usepackage{lscape}
\usepackage{pdflscape}

\def\b{\begin{eqnarray}}
\def\e{\end{eqnarray}}
\def\n{\noindent}

\begin{document}

\begin{center}

{\huge \textbf{On the Cosmological Models with \vskip.3cm  Matter Creation}}
\vskip1cm
\noindent
{\large \bf Rossen I. Ivanov and Emil M. Prodanov} \vskip.5cm
{\it School of Mathematical Sciences, Technological University Dublin, \\ City Campus, Kevin Street, Dublin, D08 NF82, Ireland,} \vskip.4cm
{\it E-Mails: rossen.ivanov@tudublin.ie, emil.prodanov@tudublin.ie} \\
\vskip1cm
\end{center}

\vskip4cm
\begin{abstract}
\n
The matter creation model of Prigogine--G\'eh\'eniau--Gunzig--Nardone is revisited in terms of a redefined creation pressure which does not lead to irreversible adiabatic evolution at constant specific entropy. With the resulting freedom to choose a particular gas process, a flat FRWL cosmological model is proposed based on three input characteristics: (i) a perfect fluid comprising of an ideal gas, (ii) a quasi-adiabatic polytropic process, and (iii) a particular rate of particle creation. Such model leads to the description of the late-time acceleration of the expanding Universe with a natural transition from decelerating to accelerating regime. Only the Friedmann equations and the laws of thermodynamics are used and no assumptions of dark energy component is made. The model also allows the explicit determination as functions of time of all variables, including the entropy, the non-conserved specific entropy and the time the accelerating phase begins. A form of correspondence with the dark energy models (quintessence, in particular) is established via the $Om$ diagnostics. Parallels with the concordance cosmological $\Lambda$CDM model for the matter-dominated epoch and the present epoch of accelerated expansion are also established via slight modifications of both models.
\end{abstract}

\vfill

\newpage

\section{Introduction}

\n
Cosmological models with adiabatic matter creation were first introduced by Prigogine {\it et al.} \cite{prigo} through the formulation of the second law of thermodynamics in the framework of general relativity. Lima {\it et al.} --- see \cite{lima0} and the references therein --- showed that cosmological models with certain rates $\Gamma$ of irreversible particle creation at the expense of gravitational energy are capable of describing the late-time acceleration of the Universe without the need to introduce dark energy, that is, such models offer an alternative to the dark energy. Adiabatic matter creation  models are fully dominated by cold dark matter particles with non-conserved number and the matter creation is assumed to happen in such way that the specific entropy is constant. However, there are no equations from which the form of $\Gamma$ can be determined. The consensus \cite{prigo, lima0, lima, zimdahl, freaza, us} is that $\Gamma$ should be considered as an input characteristic in the phenomenological description. \\
In this paper the Universe is modelled as a perfect fluid comprising of an ideal monoatomic gas containing a single type of particles with non-conserved number. This is the first of three input characteristics of the presented analysis. It is argued that there is no need at all for  adiabaticity. Very importantly, this also allows the freedom to choose the processes that will undergo in the gas which models the content of the Universe [a quasi-adiabatic polytropic process (with negative specific heat) will be chosen] --- another input characteristics of the model. The last input characteristic will be the choice $\Gamma = 3 \beta H$, where $\beta =$ const $> 0$ and $H = \dot{a}/a$ is the Hubble parameter (with $a$ being the scale factor of the Universe). \\
As the limitations imposed by the requirement of conserved specific entropy are now also fully lifted, the cosmological model with matter creation will be studied in its generality and it will shown that all model variables can be explicitly determined as functions of time. It will be demonstrated that, for specific ranges of the model parameters, a natural transition from cosmic deceleration into acceleration occurs and the acceleration redshift will be determined (depending on the model parameters). Through $Om$ diagnostic \cite{star}, the model will be found to correspond to a quintessence model (even though there is no involvement of dark energy in the matter creation models). By allowing the model parameter $\beta$ to depend on the cosmological epoch, a form of correspondence will be established with the concordance $\Lambda$CDM model for which $\Lambda$ will be allowed to vary as the inverse second power of slow cosmological time: it will be demonstrated that such models are in agreement for the matter-dominated epoch and also for the epoch of the late acceleration.

\section{The Model of Prigogine--Geheniau--Gunzig--Nardone}

\n
Prigogine {\it et al.} showed \cite{prigo} the equivalence between the energy conservation equation and the law for adiabatic and isentropic evolution of a homogeneous and isotropic Universe, that is, the energy conservation equation, $\dot{\rho} = - 3H (\rho + p),$ which stems from the covariance ($\nabla_\mu T^{\mu \nu} = 0$) of the energy-momentum tensor $T_{\mu \nu} = u_\mu u_\nu (\rho + p) - p g_{\mu \nu}$, was shown to be equivalent to an adiabatic ($\delta Q = 0$) and isentropic ($dS = 0$) thermodynamical system with evolution $T dS = \delta Q = dU + p dV = 0$. When particles are produced into the system at the expense of the gravitational field, the entropy can be no longer conserved since the particle creation enlarges the phase-space. In this case, one needs to introduce a mechanism for entropy production within such framework and the increasing entropy will necessitate a description in terms of irreversible processes: $T dS > \delta Q = 0$, hence only matter creation could be allowed, while the reverse process must be  thermodynamically forbidden. Prigogine {\it et al.} showed that, in order to achieve entropy production, one has to redefine the energy-momentum tensor $T_{\mu \nu}$ so that a supplementary pressure $\Pi$, additional to the true thermodynamical pressure $p$, is included: $T_{\mu \nu} = u_\mu u_\nu (\rho + p + \Pi) - (p + \Pi)  g_{\mu \nu}$ and hence $dU + (p + \Pi) dV = 0$. Imposing a rather limiting requirement \cite{prigo} of conserved specific entropy $\sigma = S/N$ (that is $\dot{\sigma} = 0$, while $\dot{S} > 0$), an explicit expression for this additional (``creation") pressure was determined: $\Pi = - h \Gamma / (3H)$, where $h$ is the enthalpy density (enthalpy per volume, $\mathcal{H}/V$), $\Gamma = \dot{N}/N$ is the particle creation rate. In the case of conserved specific entropy, it immediately follows that $\dot{S}/S = \dot{N}/N = \Gamma$.  Prigogine {\it et al.} illustrate their model  \cite{prigo} with $\Gamma = \alpha H^2 > 0$, where $\alpha > 0$ and regime of contraction ($H  < 0$) being also allowed (hence the square in $H$). By choosing $\Gamma = 3 \gamma H_0 + 3 \beta H$ with $0 \le \{ \gamma, \beta \} \le 1,$ Lima {\it et al.} \cite{lima0} ensured that there is a natural transition from cosmic deceleration into acceleration. Other authors \cite{lima, zimdahl, freaza,us} make different choices for $\Gamma$. \\
The form of the pressure term $\Pi$ obtained in \cite{prigo} can also be determined from the following considerations. The Gibbs equation for a system with a varying number of particles, $T dS = dU + p dV - \mu dN$, can be re-written as $T N d \sigma = dU + p dV - \mu dV - T \sigma dN = dU + p dV - \chi dN = dU + (p + \Pi) dV$, where $\chi = \mu + \sigma T$ is the specific enthalpy (enthalpy per particle, $\mathcal{H}/N$) and $\Pi = - \chi dN / dV$ with the interpretation of the expression $dN / dV$ as $\dot{N} / \dot{V} = n \Gamma / (3H)$. Owing to the fact that the specific enthalpy $\chi$ is related to the enthalpy density $h$ via $\chi = h/n$, one immediately obtains the above $\Pi = - h \Gamma / (3H)$. Given that the Gibbs law becomes
$T N d\sigma = dU + (p + \Pi) dV$ and given that one must have $dU + (p + \Pi) dV = 0$ due to the energy conservation equation, it is obvious why conservation of the specific entropy is required.

\section{The Model with Non-conserved Specific Entropy}

\n
The first input characteristic for the proposed model is the consideration of the Universe as a perfect fluid comprising of an ideal monoatomic gas with three degrees of freedom. As usual, the Universe is studied in terms of a simple thermodynamical system \cite{baz} with its volume $V$ as the single external parameter and pressure $p$ as the single generalized force associated with the single external parameter $V$. Only one type of particles with non-conserved number $N$ will be considered. The analysis will be done in terms of the thermodynamical variables $n$ (the number density, $n = N/V$) and $T$ (the temperature). Units $c = 8 \pi G = k_B = \hbar = \epsilon_0 = 1$ will be used throughout. \\
The thermic equation of state for the ideal gas is:
\b
\label{eos}
p = nT.
\e
The mean kinetic energy of a gas particle with typical rest mass $m_0$ is $(3/2)T$. If one has $N = nV$ such particles, then the internal energy of the thermodynamical system will be $U = [m_0 + (3/2) T] nV$ and, as $U = \rho/V$ on the other hand, the relationship between the energy density, the number density, and the temperature is given by the caloric equation of state $U = U(V,T)$ or
\b
\label{rho}
\rho = n \left( m_0  + \frac{3}{2} T \right).
\e
The Gibbs equation for the thermodynamical system is:
\b
\label{gibbs_mod}
T dS = dU + p dV - \mu dN,
\e
where $\mu$ is the chemical potential. \\
Adiabaticity ($\delta Q = dU + p dV - \mu dN = 0$) will not be forced upon the model. \\
The Gibbs equation can also be written as:
\b
\label{adiab}
dU + \left( p - \mu \frac{dN}{dV} - T \frac{dS}{dV}\right) dV = 0.
\e
Only time variation of the quantities will considered, thus, $dN/dV$ is interpreted as $\dot{N} / \dot{V}$, while $dS/dV$ --- as $\dot{S} / \dot{V}$. \\
Form this form of the Gibbs equation, one can identify the term $- \mu  dN/dV - T  dS/dV$ as pressure $P$, additional to the existing thermodynamical pressure $p$. The additional pressure $P = - \mu  dN/dV - T  dS/dV$ is due to the fact that neither the number of particles is conserved, nor the entropy is conserved.  Thermodynamically, one can view  the above relationship as one describing an effective adiabatic thermodynamical system with fixed number of particles, but with an additional pressure term: $dU + (p + P) dV = 0$. The extra pressure term $P$ could be referred to as ``creation--entropy pressure" and is due to the introduction of particles to the system by some mechanism and, through this, it is also due to the increase in the total entropy $S$ as the phase space enlarges. \\
As the enthalpy of the system is $\mathcal{H} = U + pV = \mu N + T S$, in terms of the energy density $\rho = U/V$; particle number density $n = N/V$; specific entropy $\sigma = S/N$; enthalpy density $h = (U + pV)/V = \rho + p$; and specific enthalpy $\chi =  (\mu N + TS)/N = \mu + T \sigma = h/n$, the Gibbs equation can be written as:
\b
\label{gibbs_cosm}
T d \sigma = p d\!\left( \frac{1}{n} \right) +  d\! \left( \frac{\rho}{n} \right) =
\frac{1}{n} (d\rho - \chi \, dn).
\e
One then has
\b
\label{cep}
P =  - \mu \, \frac{dN}{dV} - T \, \frac{dS}{dV} = - \chi \, \frac{dN}{dV} - TN \frac{d\sigma}{dV} = - \chi \, \frac{\dot{N}}{\dot{V}} - TN \frac{\dot{\sigma}}{\dot{V}}.
\e
Consider next a Friedmann--Robertson--Walker--Lema\^itre (FRWL) metric with flat spatial three-sections:
\b
ds^2 = dt^2 - a^2(t) [dr^2 + r^2 (d \theta^2 + \sin^2 \theta \, d \phi^2)],
\e
where $a(t)$ is the scale factor of the Universe. \\
The matter energy-momentum tensor $T_{\mu \nu}$, in the presence of an additional pressure term $P$ of some origin, is given by:
\b
\label{emt}
T_{\mu \nu} = (\rho + p + P) \, u_\mu \, u_\nu - (p + P) \, g_{\mu \nu} \, ,
\e
where $u^\mu$ is the flow vector satisfying $g_{\mu \nu} u^\mu u^\nu = 1$. \\
The Friedmann equations for the perfect fluid are:
\b
\dot{a}^2 & = &  \frac{1}{3} \, \rho \, a^2, \\
\label{fr2}
\ddot{a} & = & - \, \frac{1}{6} \, [\rho \, + \, 3 \,(p \, + \, P)] \, a
\e
or, in terms of the Hubble parameter $H = \dot{a}/a$:
\b
\label{h1}
H^2 & = & \frac{1}{3} \rho, \\
\label{h2}
\dot{H} & = & - \frac{1}{2} (\rho + p + P).
\e
Energy conservation means vanishing of the covariant divergence of the energy-momentum tensor: $\nabla_\mu T^{\mu \nu} = 0$. This leads to:
\b
\label{cont}
\dot{\rho} = - 3 H (\rho + p + P).
\e
Replacing $H$ by $\dot{a}/a$ and multiplying across by $a^3$ yields $(d/dt) (\rho a^3) + (p + P) (d/dt) a^3 = 0\,\,$ or $\,\, dU + (p + P)dV = 0$ --- exactly as (\ref{adiab}), if one identifies the pressure $P$ in (\ref{emt}) with the creation--entropy pressure (\ref{cep}). In other words, the reason for absorbing the entropy and particle creation terms from (\ref{gibbs_mod}) into the new pressure term $P$ is to match the second law of thermodynamics (\ref{gibbs_mod}) with $dU + (p + P)dV = 0$, which follows from the energy conservation equation $\dot{\rho} = - 3 H (\rho + p + P)$. Thus, the energy conservation equation is equivalent to the second law of thermodynamics for an effective thermodynamical adiabatic and isentropic system exhibiting an additional pressure term $P$. \\
The continuity equation for the particles of the fluid is $N^\mu_{\phantom{\mu }; \mu} = \Psi,$ where $N^\mu = n u^\mu$ is the particle
flow vector and $\Psi = n \Gamma$ is the particle production rate. Thus
\b
\label{nn}
\dot{n}  = - 3 n H + \Psi = - 3 n H + n \Gamma.
\e
Given that $V = a^3$, one has $d\sigma / dV = \dot{\sigma}/(3HV)$. Separately, $dN / dV =  d(nV) / dV = n + \dot{n}/(3H) = \Psi/(3H).$ Hence:
\b
\label{pi}
P = - \frac{1}{3H} (\chi \Psi + nT\dot{\sigma}).
\e
The Gibbs equation in the form (\ref{gibbs_cosm}) gives
\b
\label{ds}
\dot{\sigma} = \frac{1}{T n} \left( \dot{\rho} - \chi \, \dot{n} \right) = - \, \frac{1}{T n} \, (3 H P + n \chi \Gamma).
\e
Prigogine {\it et al.} \cite{prigo} consider processes of particle creation which render the specific entropy constant (i.e. $\dot{\sigma} = 0$). In such case, the creation pressure $P$ will simply be equal to the Prigogine--Geheniau--Gunzig--Nardone creation pressure $\Pi$ which corresponds to adiabatic particle production --- conserved specific entropy, but not conserved full entropy $S$ (in the case of $\dot{\sigma} = 0$, one has $\dot{S} = S \dot{N}/N$). This pressure is given by:
\b
\label{prri}
\Pi = - \frac{\chi \Psi}{3H} = - \frac{\rho + p}{n} \, \frac{\Psi}{3H} = - (\rho + p) \frac{\Gamma}{3H}
\e
and thus
\b
\label{pp}
P = \Pi - \frac{n T}{3H} \, \dot{\sigma}.
\e
In thermodynamical variables $n$ and $T$ one has:
\b
\dot{\rho}(n, T) = \biggl(\frac{\partial \rho}{\partial n}\biggr)_T \dot{n} + \biggl(\frac{\partial \rho}{\partial T}\biggr)_n \dot{T}.
\e
Substituting the energy conservation equation (\ref{cont}) and the particle conservation equation (\ref{nn}), using the thermodynamic identity
\b
\label{tdi}
h = \rho + p = T \biggl( \frac{\partial p}{\partial T}\biggr)_n + n \biggl( \frac{\partial \rho}{\partial n}\biggr)_T
\e
and also (\ref{pp}), leads to the following temperature evolution law:
\b
\label{temp}
\frac{\dot{T}}{T} =
\biggl(\frac{\partial p}{\partial \rho}\biggr)_{\!\!n}  \frac{\dot{n}}{n} + \frac{n \dot{\sigma}}{\left(\!\frac{\partial \rho}{\partial T}\!\right)_{\!n}}.
\e
In the absence of particle creation and with conserved specific entropy, this reduces to $\dot{T}/T = - 3H (\partial p / \partial \rho)_n$. \\
The entropy of the system will depend on the type of gas used for the model of the Universe, on the type of the process involved, and also on the particle creation rate $\Gamma$. All these are input characteristics of the model as they cannot be determined from any equations. Amongst the so far  presented equations (or, in other words, the laws of thermodynamics and general relativity), there is not one from which one can determine the entropy.  \\
Calv\~ao, Lima, and Waga \cite{lima} introduce, through an ansatz, the following form of the creation pressure:
\b
\label{ansatz}
P = - \alpha \frac{\Psi}{3H}
\e
where $\alpha$ is positive. \\
This is nothing else but an additional constraint on $P$, since $P$ was already determined when the second law of thermodynamics was cast into the form $dU + (p + P)dV = 0$ which, in turn, stems from the energy conservation equation (\ref{cont}). Equating $P$ from this ansatz to $P$ determined in (\ref{pi}), gives:
\b
\label{se}
\dot{\sigma} = \frac{\Gamma}{T} ( \alpha - \chi).
\e
This relationship is also obtained in \cite{lima} and it is obvious that when $\alpha = \chi = (\rho + p)/n$, then the creation pressure is equal to the Prigogine--Geheniau--Gunzig--Nardone creation pressure $\Pi$, the specific entropy is conserved, and the adiabatic picture of Prigogine {\it et al.} \cite{prigo} applies. \\
However, in the case of entropy production, $\alpha$ in (\ref{ansatz}) remains undetermined and the resulting equation (\ref{se}) leaves the entropy production in turn undetermined. \\
To determine the produced entropy, as already mentioned, one needs to commit to a particular type of gas for the model of the Universe, a particular process, and a particular particle creation rate. \\
Since $(\partial p / \partial \rho)_n = (\partial p / \partial T)_n / (\partial \rho / \partial T)_n$, for an ideal gas the temperature law (\ref{temp}) becomes
\b
\frac{\dot{T}}{T} = \frac{2}{3}\,\left( \frac{\dot{n}}{n} + \dot{\sigma} \right).
\e
This integrates to give
\b
\label{fi}
T = \tau \, n^{\frac{2}{3}} \, e^{\frac{2\sigma}{3}}
\quad \mbox{ or } \quad \sigma = \frac{3}{2} \, \ln \!\left( \frac{T}{\tau} n^{-\frac{2}{3}} \right)\!,
\e
where $\tau$ is an integration constant (temperature scale). To determine $\tau$, consider the following. If a particle has $g$ internal degrees of freedom, then the density of states in the phase space is given by $g (2 \pi)^{-3} f(p)$, where the distribution function
\b
f(p) = \left[ e^{\frac{E(p) - \mu}{T}} \pm 1 \right]^{-1}.
\e
for a system of particles in equilibrium is given by the Fermi-Dirac distribution functions for fermions (positive sign) or the Bose-Einstein distribution function for bosons (negative sign). Here $\mu = \chi - \sigma T$ is the chemical potential and $E(p) = \sqrt{m_0^2 + p^2} = m_0 + p^2/2m_0$ is the energy of a particle of rest mass $m_0$ and momentum $p$. \\
The particle number density $n$ is obtained after integrating $g (2 \pi)^{-3} f(p)$ over the momentum
\b
\label{df}
n & = & \frac{g}{(2 \pi)^{3}} \int d^3p \, f(p) = g \left( \frac{m_0 T}{2 \pi} \right)^{\!\!\frac{3}{2}}
e^{ -\frac{m_0 - \mu}{T}} \nonumber \\
& = &   \left( \frac{m_0 g^{\frac{2}{3}} }{2 \pi} \right)^{\!\!\frac{3}{2}} T^{\frac{3}{2}}
e^{ -\frac{m_0 - \chi}{T}}  e^{-\sigma} =  \left( \frac{m_0 g^{\frac{2}{3}} e^{\frac{5}{3}}}{2 \pi} \right)^{\!\!\frac{3}{2}} T^{\frac{3}{2}}
e^{-\sigma}. \nonumber \\
\e
since $(\chi - m_0)/T = [(\rho + p)/n - m_0]/T = 5/2$. \\
Thus
\b
\label{sub}
\sigma = \frac{3}{2} \, \ln \! \left(\frac{m_0 g^{\frac{2}{3}}e^{\frac{5}{3}}}{2\pi} T n^{-\frac{2}{3}} \right)
\e
and hence
\b
\tau = \frac{2 \pi}{m_0 g^{\frac{2}{3}}e^{\frac{5}{3}}}
\e
--- see also equation (46.1a) in \cite{landau}. \\
Two further equations are needed in order to determine how the two independent thermodynamical variables $n$ and $T$ depend on time and from this --- how all other variables of the model depend on time. One of these equations is already available --- this is the particle conservation equation, $\dot{n}/n = - 3H + \Gamma$. To avail of this equation, a specific choice of the particle creation rate $\Gamma$ has to be made. Following \cite{lima0, lima, zimdahl, freaza}, the particle creation rate will be taken as $\Gamma = 3 \beta H$ with $\beta > 0$ and $H > 0$ (in view of the current state of the Universe, only regime of expansion will be considered: $H = \dot{a}/a > 0$) and this is the second input characteristic of the model. Note that, because of $\dot{N}/N = \Gamma$, the positivity of $\Gamma$ leads to $\dot{N} > 0$, i.e. only a particle creation process is considered. But this is not the only possibility --- see \cite{zimdahl, freaza, us}. \\
To derive the needed equation for the evolution of the temperature, a third (and final) input characteristic is needed --- the type of process which the ideal gas undergoes. With the freedom to chose, a polytropic process $T dS = \delta Q = N c dT$, where $c = $ const is the specific heat of the expanding Universe. To ensure increasing entropy in the regime of decreasing temperature, one must have $c < 0$.  Such polytropes are called quasi-adiabatic processes \cite{gr}. \\
Polytropes have very wide applications in astrophysics and the related fields --- see the extensive monograph \cite{horedt}. Polytropic gas models of dark energy provide alternative explanation of the accelerated expansion of the Universe --- see the review \cite{gr} and the references therein. These models follow the steps of the Cardassian expansion cosmological models \cite{card} for which the right-hand side of the Friedmann equation $H^2 = (1/3)\rho$ is modified to involve an additional, polytropic, term: $H^2 = (1/3) \rho + B \rho^n$, where $B$ is some constant and $n < 2/3$ in order to achieve accelerated expansion. The authors of these models also allow a more general function $f(\rho)$ to be added to the modified Friedmann equation and this is referred to as generalized Cardassian model. The Cardassian models fit with both quintessence and phantom cosmology. A phenomenological model in which the pressure density of (phantom) dark energy is given by the generalized $p = -\rho - f(\rho)$ was further investigated by \cite{noj} in the context of the study of future singularities; see \cite{stef} for the study of the future singularities in the case of $f(\rho) = B \rho^n$ and see also \cite{further} and the references therein for further developments, including interacting dark energy. \\
For the polytropic process $T dS = \delta Q = N c dT$ (with $c < 0$), the first law of thermodynamics, $N c dT = dU + pdV - \mu dN$, becomes:
\b
( c - \frac{3}{2}) \frac{dT}{T} = \frac{dV}{V} + \left( \frac{3}{2} + \frac{m_0 - \mu}{T} \right) \frac{dN}{N}.
\e
In the regime of low particle number densities $n$, one has $\rho \approx m_0 n$. Thus, $p = n T \approx (\tau^*/m_0^{5/3}) \rho^{5/3} > 0$. This corresponds to a polytropic process in a monoatomic ideal gas with three degrees of freedom: $p = ($const$) \rho^{(\alpha + 1)/\alpha}$ with $\alpha = 3/2$, giving a polytropic index of 5/3, equal to the heat capacity ratio of the gas. \\
Given that $\sigma = (\chi - \mu)/T = [m_0 + (5/2)T - \mu]/T$ and also using $(dN)/N = \Gamma dt$ and $(dV)/V = (da^3)/a^3 = 3Hdt$ gives:
\b
\left( c - \frac{3}{2} \right) \frac{\dot{T}}{T} = 3H - \Gamma + \Gamma \sigma.
\e
This equation and the particle conservation equation
\b
\frac{\dot{n}}{n} = - 3H + \Gamma
\e
form a two-dimensional autonomous dynamical system for the two thermodynamical variables $n(t)$ and $T(t)$. \\
Substituting $\rho$ from the caloric equation of state (\ref{rho}) into the Friedmann equation $H = + \sqrt{\rho/3} > 0$ yields $ H = + (\sqrt{3}/3) \sqrt{m_0 n + (3/2) nT} > 0$  and the dynamical system in the case of $\Gamma = 3 \beta H$ can be written as:
\b
\label{ds1a}
\!\!\!\!\!\!\frac{\dot{T}}{T} & = & \frac{\sqrt{3}}{c - \frac{3}{2}}  \sqrt{ n m_0 + \frac{3}{2} n T }  \left[ 1 - \beta + \frac{3}{2} \beta \ln \!\left( \frac{T}{\tau} n^{-\frac{2}{3}} \right) \!\right]\!\!, \\
\label{ds2a}
\!\!\!\!\!\!\frac{\dot{n}}{n} & = & \sqrt{3} \, (\beta - 1) \,\sqrt{ n m_0 + \frac{3}{2} n T }.
\e
This dynamical system is integrable. To see this, introduce variables $x = \ln n$ and $y = \ln T$ and divide the two equations to get:
\b
\label{niama}
\frac{dy}{dx} = \frac{1}{\kappa}( 1 - \beta + \beta \sigma) = \frac{1}{\kappa} \left(\alpha + \frac{3}{2}\beta y - \beta x \right),
\e
where $\kappa = (c - 3/2)(\beta - 1)$ and $\alpha = 1 - \beta - (3/2) \beta \ln \tau$. The solution of this equation is
\b
\ln \left( \frac{T}{\tau} \, n^{-\frac{2}{3}} \right) = D \, n^{\frac{3\beta}{2\kappa}} \,\, + \,\, \frac{4c(\beta - 1)}{9\beta},
\e
where $D$ is an integration constant. The value of $D$, depends on the prescribed initial conditions $T_0 = T(n_0)$. \\
In view of (\ref{fi}):
\b
\label{tuuk}
\sigma =  \frac{3}{2} \, D \, n^{\frac{3\beta}{2\kappa}} \,\, + \,\, \frac{2c(\beta - 1)}{3\beta}.
\e
If the integration constant $D$ is zero, then the specific entropy is constant. Thus $D = 0$ corresponds to adiabatic particle creation. For a positive specific entropy, one must have $D > 0$ [the constant $2c(\beta - 1)/(3\beta)$ will turn out to be also positive]. \\
The temperature $T$ as function $n$ is therefore given by
\b
\label{tem}
T(n) = \tau^* \, n^{\frac{2}{3}} \, e^{Dn^{\frac{3\beta}{2\kappa}}},
\e
where $\tau^* = \tau \exp [4c(\beta - 1)/(9\beta)] = $ const. \\
Substituting this temperature law into equation (34) yields:
\b
\label{enn}
\dot{n} = \sqrt{3} \, (\beta - 1) \, n^{\frac{3}{2}} \, \sqrt{m_0 \, + \, \frac{3}{2} \, \tau^* \, n^{\frac{2}{3}} \, e^{D n^{\frac{3 \beta}{2 \kappa}}}}.
\e
In view of the transcendental character of this equation, the complete evolution of $n(t)$ cannot be given explicitly in terms of elementary functions. Proper phase-plane analysis, for example in the $n$--$H$ plane, would reveal qualitatively all features of the bahaviour of the system. \\
Physically, it makes sense to have $T \to 0$ when $n \to 0$. Thus, given that $\beta > 0$, one has to have $\kappa > 0$, which, in turn, leads to  $0 < \beta < 1$ since $c < 0$. Having $0 < \beta < 1$ also avoids a model in which, owing to (\ref{ds2a}), $n$ increases with time (and, together with it, $T$), i.e.  the parameter $\beta$ in the particle creation term $\Gamma = 3 \beta H$ is restricted. \\
Equipped with $T(n)$ from (\ref{tem}), equation (\ref{ds2a}) becomes an equation for $n(t)$ in separate variables and can be integrated (albeit not analytically). In view of this, it is best to study the dynamical system (\ref{ds1a})--(\ref{ds2a}) numerically in order to determine $T(t)$ and $n(t)$. \\
For small $n$, the dynamical equation (\ref{ds2a}) in leading order is:
\b
\dot{n} \approx \sqrt{3m_0} (\beta - 1) n^{\frac{3}{2}}.
\e
After integration one gets:
\b
\label{using}
n(t) = \frac{1}{\left[ \sqrt{\frac{1}{n_0}} - \frac{\sqrt{3m_0}}{2} (\beta - 1) (t - t_0) \right]^2},
\e
where $n_0 = n(t_0)$. \\
As time increases ($t \to \infty$), the number density behaves so that
\b
\sqrt{n(t)} \approx \frac{2}{\sqrt{3m_0} \, \vert \beta - 1 \vert \, t}.
\e
Thus
\b
\frac{\dot{a}}{a} = \frac{d}{dt} \, \ln a = H \approx \sqrt{\frac{m_0}{3}} \, \sqrt{n} \approx \frac{2}{3 \, \vert \beta - 1 \vert \, t}.
\e
For the scale factor one gets:
\b
\label{evo}
a(t) & = & a_0 \, t^{\,\frac{2}{3 \, \vert \beta - 1 \vert}}.
\e
The evolution of the scale factor $a(t)$ in the general case can be obtained after integration of
\b
\frac{\dot{a}}{a} =  \frac{\sqrt{3}}{3} \, \sqrt{m_0 n \, + \, \frac{3}{2} \, \tau^* \, n^{\frac{5}{3}} \, e^{D n^{\frac{3 \beta}{2 \kappa}}}},
\e
where $n = n(t)$ is the solution of (\ref{enn}). \\
When there is no particle creation (i.e. $\beta = 0$), the time-dependence of the scale factor is the well known $a(t) \sim t^{2/3}.$ Accelerated expansion ($\dot{a} > 0$ and $\ddot{a} > 0$) is achieved for $1/3 < \beta < 1$. Smaller values of $\beta$ (between 0 and 1/3) correspond to particle creation that does not generate enough particles to trigger acceleration of the expansion. In summary, as the particle creation rate is proportional to $H$ (with $H \sim 1/t$), for $0 < \Gamma < H$, the Universe is expanding, without acceleration, towards $n \to 0, \,\, T \to 0$. With a higher particle creation rate, namely for $H < \Gamma < 3H$, the expansion is accelerating --- towards $n \to 0, \,\, T \to 0$ again. Finally, for $\Gamma > 3H$ (corresponding to $\beta > 1$), the particle creation rate is so high that $n$ increases with time.  \\
Due to (\ref{tuuk}), one has:
\b
\label{nega}
\!\!\!\dot{\sigma} = \frac{9 \beta}{4 \kappa} D n^{\frac{3\beta}{2\kappa}-1} \, \dot{n} = \frac{27 D \beta (\beta - 1)}{4 \kappa} \,\, H \, n^{\frac{3\beta}{2\kappa}} < 0
\e
in view of $\beta < 1$, i.e. the specific entropy $\sigma = S/N$ decreases with time. This is not a contradiction to the second law of thermodynamics, it simply means that $N$ grows faster than $S$. The full entropy $S$ increases with time as can be easily seen: one has $\dot{S} = (d/dt)(\sigma a^3 n) = N (\dot{\sigma} + \Gamma \sigma)$ and thus
\b
\!\!\!\dot{S} = \frac{3NH}{2} \left[ \frac{3\beta D}{\kappa}(c - 1)(\beta-1) n^{\frac{3\beta}{2\kappa}} + \frac{4}{3} c (\beta - 1) \right]
\e
which is positive.

\section{Accelerated Expansion and the Acceleration Redshift}

The accelerated expansion is not an ever-present feature of the model. To understand when (at what redshift) the acceleration becomes positive, consider the Friedmann equation (\ref{fr2}) and note that $\rho + 3(p+P) < 0$ is needed for $\ddot{a} > 0$ (acceleration of the expansion).
\b
- \, 6 \frac{\ddot{a}}{a} \hskip-.3cm & = & \hskip-.3cm  \rho + 3(p + P) = \rho + 3 \left(p + \Pi - \frac{n T}{3H} \dot{\sigma}\right) = \rho + 3\left[ p - \beta (\rho + p) - \frac{n T}{3H} \dot{\sigma}\right], \nonumber \\
\e
as, in view of (\ref{prri}), $\Pi = - (\rho + p) \Gamma/(3H) = - \beta (\rho + p)$ since $\Gamma = 3 \beta H$. Further, in light of (\ref{nega}), one has $\dot{\sigma} = 27 D \beta (\beta - 1)/(4 \kappa)\,\, H \, n^{\frac{3\beta}{2\kappa}}$. Using equation (\ref{rho}) for the relationship between $\rho$ and $n$ and the equation of state (\ref{eos}) yields:
\b
\label{46}
- \, 6 \frac{\ddot{a}}{a} \,\, = \,\, (1 - 3 \beta)\,(m_0 n + \frac{3}{2} nT) \, + \, 3(1 - \beta) nT  \, - \, \frac{9 D \beta (\beta - 1)}{4 \kappa} \, T \, n^{\frac{3\beta}{2\kappa} + 1}.
\e
Substituting the temperature law (\ref{tem}), one finally gets the acceleration in terms of $n$ only:
\b
- \, 6 \frac{\ddot{a}}{a} \hskip-.2cm & = & \hskip-.2cm
(1 - 3 \beta) m_0 n -  \frac{\tau^*}{2} (15\beta - 9)  n^{\frac{5}{3}}  e^{Dn^{\frac{3\beta}{2\kappa}}}
+  \frac{9 \tau^*  D \beta (1 - \beta)}{4 \kappa} n^{\frac{3\beta}{2\kappa} + \frac{5}{3}}  e^{Dn^{\frac{3\beta}{2\kappa}}}. \nonumber \\
\e
Thus, there is some value of $n$, say $n_c$, which depends on all model parameters, $n_c = n_c(m_0, D, \epsilon, c)$, and for which the expression in the brackets changes sign (the last two terms increase monotonically with $n$). Therefore, the transition from decelerated to accelerated behaviour occurs when $n$ drops to $n_c$. Although the form of $n_c$ is not available explicitly in terms of elementary functions, it is clear that such value exists for small $n$.  \\
As already seen, values of $\beta$ greater than 1/3 lead to power law accelerated expansion and there is no acceleration if $0 < \beta \le 1/3$. Also, $\beta$ must be smaller than 1 so that $T \to 0$ when $n \to 0$. To investigate the critical value $\beta_c = 1/3$, consider $\beta = 1/3 + \epsilon$, where $0 < \epsilon \ll 1$. Then (\ref{46}) becomes:
\b
\label{qqqq}
- \, 6 \frac{\ddot{a}}{a} &  = & n \, T(n) \left( -\frac{3 m_0 \epsilon}{T(n)} + 2 + \frac{D}{2 \kappa} \, n^{\frac{1}{2\kappa}}  \right) \mbox{ with  } T(n) = \tau^* \, n^{\frac{2}{3}} \, e^{Dn^{\frac{1}{2 \kappa}}}. \nonumber \\
\e
Note that $\kappa = 1 - 2c/3$ when $\beta = 1/3 + \epsilon$. \\
For values of $\beta$ well above $\beta_c$ (but still $\beta < 1$), accelerated expansion is always present within this non-relativistic model. \\
To express the model quantities as functions of the redshift $z$, instead of time $t$, introduce the cosmological redshift $z$ for a flat FRWL space-time via
\b
1 + z = \frac{a_0}{a}.
\e
Thus, $dz = - (a_0 \dot{a}/a^2) dt = - (1 + z) H dt$. One immediately gets:
\b
\frac{dH}{dt} & = & \frac{dH}{dz} \frac{dz}{dt} = - (1 + z) H \frac{dH}{dz} = - \frac{1}{2} (1 + z) \frac{dH^2}{dz},  \\
\frac{dn}{dt} & = & \frac{dn}{dz} \frac{dz}{dt} = - (1 + z) H \frac{dn}{dz}.
\e
Using the particle conservation equation $\dot{n} = 3 (\beta - 1) H$ on the left-hand side of the latter yields:
\b
\frac{dn}{dz} = - \frac{3(\beta - 1)}{1 + z}.
\e
This integrates easily to give
\b
n(z) = n_0 (1 + z)^{3(1 - \beta)}.
\e
The deceleration parameter $q$, given by:
\b
q = - \frac{a \ddot{a}}{\dot{a}^2} = - \frac{\ddot{a}}{a H^2},
\e
with the help of (\ref{qqqq}), i.e. for small $n$ and $\beta = 1/3 + \epsilon$, can be written as
\b
q = \frac{n \, T(n)}{6 H^2} \, \left[ 2 - \frac{3 m_0 \epsilon}{T(n)} +\frac{D}{2\kappa} \, n^{\frac{1}{2 \kappa}} \right].
\e
For small $n$, one has $T(n) \approx \tau^* n^{2/3}$. In terms of the redshift $z$, one immediately gets $T(z) \approx \tau^* n_0^{2/3} [(1+z)^{3(1-\beta)}]^{2/3}$. When $\beta = 1/3 + \epsilon$, one further gets $T(z) \approx \tau^* n_0^{2/3} (1+z)^{4/3}$. \\
Introduce, for simplicity, the scaled deceleration parameter $\tilde{q}(z)$ given by $\tilde{q}(z) = 6 H^2(z) \, n^{-1}(z) \, T^{-1}(z) q(z)$. The zeros of $\tilde{q}(z)$ and $q(z)$ coincide. Then, for $\tilde{q}(z)$ one has:
\b
\tilde{q}(z) = 2 - \frac{k}{(1+z)^{\frac{4}{3}}} + \mbox{small terms},
\e
where $k = (3 m_0 \epsilon)/(\tau^* n_0^{2/3}) = $ const.

\begin{figure}[!ht]
\centering
{\includegraphics[height=5cm, width=0.48\textwidth]{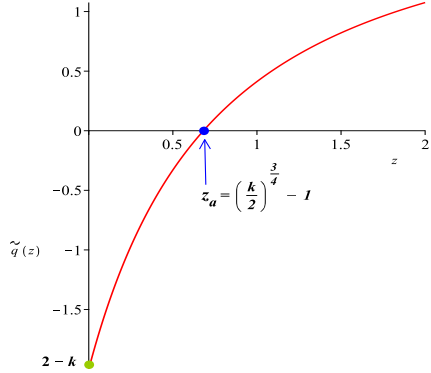}}
\caption{\footnotesize{The scaled deceleration parameter $\tilde{q} = ( 6 H^2 ) / \left( nT \right) \, q$ (plotted with $k = 4$ which yields acceleration redshift $z_a = 0.68$).}}
\end{figure}
\n
Clearly, the graph of $\tilde{q}(z)$ intercepts the ordinate at $2 - k$, while the acceleration redshift is $z_a = (k/2)^{3/4} - 1 = [(3 m_0 \epsilon)/(2 \tau^* n_0^{2/3}) ]^{3/4} - 1$ --- see Figure 1. The parameters $m_0, \epsilon, \tau^*,$ and $n_0$ could be chosen in such way that $\tilde{q}(z) > 0$ for all $z$, namely, for $\tilde{q}(z) > 0$ for all $z$, one needs $k = (3 m_0 \epsilon)/(\tau^* n_0^{2/3}) < 2$. \\
To relate to the experimental limits $0.4 \le z_a \le 0.8$ \cite{star}, the parameters $m_0, \epsilon, \tau^*,$ and $n_0$ should be such that $1.4 \le [(3 m_0 \epsilon)/(2 \tau^* n_0^{2/3}) ]^{3/4} \le 1.8$. This puts limits on $k$ and hence one can find the limits on $\epsilon = (k \, \tau^* n_0^{2/3})/(3 m_0)$. \\
The functional dependence of the scaled deceleration parameter $\tilde{q}$ on the redshift $z$ is in agreement with the one emerging from $\bar{q}$ diagnostic of Union supernovae data, reported in \cite{star} --- compare Figure 1 to Figure 7 in \cite{star}.

\section{Link to Dark Energy Models}

Dark energy models are based on a component with equation of state $p = \omega \rho$ and the Friedmann equation in these models, namely $\ddot{a}/a = - (1/6) (\rho + 3p) = - (1/6) (1 + 3 \omega) \rho$, dictates that for cosmic acceleration ($\ddot{a} > 0$), one has to have $\omega < - 1/3$ and thus, negative pressure. This also results in the violation of the strong energy condition: $\rho + p \ge 0$ and $\rho + 3p \ge 0$. When $\omega = - 1$, the resulting model is the standard (concordance) cosmological model $\Lambda$CDM. Models with $-1 < \omega < -1/3$ are called quintessence models, while those with $\omega < -1$ are called phantom cosmological models. The latter violate all four energy conditions. Recently in \cite{star}, a diagnostic test called $Om$ has been proposed, which is constructed from the Hubble parameter $H = \dot{a}/a$ with the latter determined directly from observational data. This diagnostic provides a {\it null test} of the $\Lambda$CDM hypothesis and allows for the differentiation between various dark energy models \cite{star}. If the $Om$ diagnostics, as a function of the redshift $z$, that is $Om(z)$, is constant (equal to the value of the matter density $\Omega_{0m}$) for all $z$, then the model in question is the concordance $\Lambda$CDM model ($\omega = -1$). For dark energy models with dynamical equation of state, a positive slope of $Om(z)$ suggests a phantom cosmological model ($\omega < - 1$), while a negative slope corresponds to a quintessence model ($- 1 < \omega < - 1/3$) \cite{star}. The $Om$ diagnostics provides such distinction between various dark energy models both with and without reference to the value of the matter density $\Omega_{0m}$, thus having the ability to  avoid a potential source of significant uncertainty in the cosmological reconstruction \cite{star}. \\
The $Om$ diagnostic is introduced in the following manner \cite{star}:
\b
Om(x) = \frac{h^2(x) - 1}{x^3 - 1},
\e
where $x = 1 + z$ and $h(x) = H(x)/H_0$. \\
For the $\Lambda$CDM model [$\omega(z) = $ const] one has \cite{star}:
\b
h^2(x) = \Omega_{0m} x^3 + (1 - \Omega_{0m}) x^{\alpha} \quad \mbox{with} \quad \alpha = 3 (1 + \omega).
\e
Thus \cite{star}:
\b
Om(x) = \Omega_{0m} + (1 - \Omega_{0m}) \, \frac{x^{\alpha} - 1}{x^3 - 1}.
\e
If the dark energy is modelled by the cosmological constant [$\omega(z) = $ const $ = - 1$], i.e. for the $\Lambda$CDM model, one has $\alpha = 0$ and hence $Om(x) = \Omega_{0m}$. \\
Dynamical dark energy models [having $\omega = \omega(z)$], as opposed to the standard cosmological $\Lambda$CDM model, can also explain the observational data with which the $\Lambda$CDM model is in excellent agreement --- see \cite{star} and the references therein. The following parametric ansatz for $\omega(z)$ is made \cite{star, cpl}:
\b
\omega(z) = \omega_0 + \omega_1 \, \frac{z}{1+z}.
\e
Then, if $Om(x) > \Omega_{0m}$, then the model describes quintessence ($\alpha > 0$), while if $Om(x) < \Omega_{0m}$, one has a phantom cosmological model ($\alpha  < 0$), \cite{star}. The graph of the former has a negative slope, while that of the latter has a positive slope --- see the figures in \cite{star}. \\
The cosmological models with particle creation do not describe dark energy: such models are alternative to the dark energy models. In order to establish a form of correspondence between the proposed model with particle creation and the dark energy models, $Om$ diagnostics will be applied to the particle creation model. One has
\b
Om(x) = \frac{\left[ \frac{H(n)}{H_0(n_0)} \right]^2 - 1}{x^3 - 1} = \frac{\frac{m_0 n + \frac{3}{2} n T(n)}{m_0 n_0 + \frac{3}{2} n_0 T(n_0)} - 1}{x^3 - 1}.
\e
To study the $Om$ diagnostic for values in the leading order of the particle number density $n$ (namely, disregarding the contributions of the non-conserved specific entropy $\sigma$), expand the temperature evolution law (\ref{tem}):
\b
T(n) = \tau^* \, n^{\frac{2}{3}} \, e^{Dn^{\frac{3\beta}{2\kappa}}} = \tau^* n^{\frac{2}{3}} + \ldots
\e
and substitute in the above to get
\b
Om(x) & \hskip-0.65cm = \hskip-0.65cm & \frac{\frac{n}{n_0} \left(1 + \frac{3 \tau^*}{2 m_0} n^{\frac{2}{3}} \right) \!\! \left(1 + \frac{3 \tau^*}{2 m_0} n_0^{\frac{2}{3}} \right)^{-1} \!\! - 1}{x^3 - 1} =  \frac{\frac{n}{n_0} \Biggl[ 1 + \xi \! \left[ \left( \frac{n}{n_0} \right)
^{\frac{2}{3}} \!\! - 1 \right] \Biggr] \!\! - 1}{x^3 - 1}, \nonumber \\
\e
where $\xi = (3 \tau^* n_0^{2/3})/(2m_0) = $ const. \\
Given that $n/n_0 = x^{3(1-\beta)}$, the application of the $Om$ diagnostic for values of the parameter $\beta$ near the critical value $\beta_c = 1/3$, i.e. $\beta = 1/3 + \epsilon$ with $0 < \epsilon \ll 1$, gives
\b
\label{ooo}
Om(x) = \frac{x + 1}{x^2 + x + 1} + \xi \, \frac{x^2 \, \left( x^{\frac{1}{3}} + 1 \right)\, \left( x^{\frac{2}{3}} + 1 \right)}{\left( x^{\frac{2}{3}} +  x^{\frac{1}{3}} + 1 \right)\, \left( x^2 + x + 1 \right)}.
\e
In the numerator of $\xi$ one has $T_0$ in the leading order of $n_0$ and within the range of validity of the model, $T/m_0 \ll 1$ for all $T$, including $T_0$. Therefore $\xi \approx 10^{-9}$ or less  for the present epoch as, more precisely, the characteristic particles of the model are of rest mass of about  $0.5$ MeV or more, that is, $10^9$ K or more. Even the highest possible redshift cannot compensate the smallness of the second term in (\ref{ooo}) and for that reason this term should be neglected. Thus
\b
Om(x) = \frac{x + 1}{x^2 + x + 1}.
\e
With the increase of $x$, this function tends monotonically from 1 (at $x = 0$) to zero. As the slope is always negative, the particle creation model bears the hallmarks of a quintessence model.

\section{Correspondence with the $\Lambda$CDM model}

As the model is applicable to the matter-domination epoch and the present epoch of accelerated expansion, to establish a relation to the concordance $\Lambda$CDM cosmological model for these two stages of the cosmological evolution, assume that the parameter $\beta$ could vary from one epoch to another. This is, in fact, a modification of the particle creation model making it differ from what was associated with quintessence through the $Om$ diagnostics in the previous section. \\
The matter-domination laws correspond to $\beta = \beta_{md} \ll 1$, which, using (\ref{evo}), leads to $a(t) \sim t^{2/3}$. \\
For the present epoch of accelerated expansion, observational data is fitted by the $\Lambda$CDM model\footnote{Recently, some tension has been reported between different high redshift observations and the $\Lambda$CDM model  --- see \cite{more}.} with $a(t) = a_0 \exp \!\left( \sqrt{\Lambda / 3} t \right)$, where $\Lambda$ is the cosmological constant. Thus, $H = \dot{a}/a = \sqrt{\Lambda/3}$. To draw a parallel to the presented model, the substitution of the latter into the dynamical equation $\dot{n} / n = - 3 (1 - \beta_{pe}) H$ yields $\dot{n} / n = - 3 (1 - \beta_{pe}) \sqrt{\Lambda/3}$. On the other hand, from (34), for low temperatures, one has $\dot{n} / n = - \sqrt{3} (1 - \beta_{pe}) \sqrt{m_0 n}$. Comparison of these two gives $\Lambda \approx m_0 n$. Using equation (\ref{using}), one gets
\b
\Lambda \approx \frac{m_0 \, n_0^*}{\left[ 1 + \frac{\sqrt{3 m_0 n_0^*}}{2} \, (1 - \beta_{pe}) \, (t - t_0^*) \right]^2},
\e
where $m_0 n_0^* \approx \Lambda(t_0^*)$ is the energy density in the early stages ($t_0^*$) of the present epoch of accelerated expansion. The time-dependence in the above would be very weak provided that $\beta_{pe} = 1 - \tilde{\epsilon}$ with $0 < \tilde{\epsilon} \ll 1$. In this way, $\Lambda$ would depend on the ``slow" time $t_s = \tilde{\epsilon} (t - t_0^*)$ and would be a constant in the limit of $\tilde{\epsilon} \to 0$. Dependence of $\Lambda$  on time is not a new has been widely discussed.  Weinberg argues \cite{weinberg} that if the cosmological constant is small now, it was not necessarily always small. Dirac's large number hypothesis \cite{dirac} leads to a cosmology where $\Lambda$ varies very slowly with cosmological time. Dirac argues that $H \sim t^{-1}$. Lima and Carvalho \cite{lc} propose $\Lambda \sim H^2$, thus $\Lambda \sim t^{-2}$ --- in agreement with the above.

\section{Conclusions}

The aim of this paper is to prove the concept that the specific entropy $\sigma$ in cosmological models with matter creation does not need to be constant. This new feature is studied in the presented model, together with the viable cosmological consequences it leads to. On one hand, the significant physical restriction presented by the requirement $\sigma = $ const has been lifted. On the other hand, new horizons are revealed, in particular, it is exactly the non-conservation of the specific entropy that allows one to uncover the possible transition from decelerated phase to the current accelerated phase for certain values of the model parameters, that is, models with $\beta$ close to the critical value $1/3$ have the remarkable property that the transition from deceleration to accelerated expansion (hence between cosmological epochs) happens  naturally with the decreasing of the particle number density $n$ (or the temperature $T$). \\
The presented analysis is in line with the existing particle creation models with conserved specific entropy, whose results can be reproduced by setting $D = 0$; with the cosmological models without particle creation (reproduced by setting $\beta = 0$); and with the models with adiabatic non-accelerated expansion of the Universe (by taking $c = 0$). \\
The $Om$ diagnostic allows the association of the model with a quintessence model. Additionally, the presented model with $\beta$ allowed to vary with the epoch qualitatively matches the behavior of the scale factor $a(t)$ from the standard cosmological model for the matter domination epoch [for early times: $a(t) \sim t^{2/3}$] and for the epoch of the late acceleration [for late times: $a(t) \sim e^{t/t_{\Lambda}}$]. This analogous $\Lambda$CDM model has cosmological constant varying with a very slow cosmological time.

\end{document}